\input{aipcheck.tex}
\documentclass[final]{aipproc}
\layoutstyle{8x11single}
\def\be{\begin{equation}}
\def\ee{\end{equation}}
\def\bea{\begin{eqnarray}}
\def\eea{\end{eqnarray}}
\def\ba{\begin{array}}
\def\ea{\end{array}}
\usepackage{bm}

\begin{document}
\title[]{Chiral symmetry breaking and electromagnetic structure of the nucleon}
\classification{13.40.Gp, 12.39.Fe} \keywords{Electromagnetic form
factors, chiral symmetry breaking, charge radii, chiral
constituent quark model}
\author{Harleen Dahiya}{address={Department of Physics, Dr. B.R. Ambedkar National Institute of Technology,
Jalandhar, Punjab-144 011, India.}}
\author{Neetika Sharma}{address={Department of Physics, Dr. B.R. Ambedkar National Institute of Technology,
Jalandhar, Punjab-144 011, India.}}

\date{\today}

\begin{abstract}

The electromagnetic form factors are the most fundamental
quantities to describe the internal structure of the nucleon and
are related to the charge radii of the baryons. We have calculated
the charge radii of octet baryons in the framework of chiral
constituent quark model with the inclusion of the spin-spin
generated configuration mixing. This model is quite successful in
predicting the low energy hadron matrix elements. The results
obtained in the case of charge radii are also comparable to the
latest experimental studies and show improvement over some
theoretical interpretations.
\end{abstract}

\maketitle

\section{Introduction}

The electromagnetic form factors are the fundamental quantities of
theoretical and experimental interest to investigate the internal
structure of nucleon. The knowledge of internal structure of
nucleon in terms of quark and gluon degrees of freedom in QCD
provide a basis for understanding more complex, strongly
interacting matter. Recently, a wide variety of accurately
measured data have been accumulated for the static properties of
baryons, for example, masses, electromagnetic moments, charge
radii, and low energy dynamical properties such as scattering
lengths and decay rates etc.. The charge radii and magnetic
moments, as measured for the distribution of charge and
magnetization, represent important observables in hadronic physics
as they lie in the nonperturbative range of QCD and give valuable
information on the internal structure of hadrons. While QCD is
accepted as the fundamental theory of strong interactions, it
cannot be solved accurately in the nonperturbative regime. A
coherent understanding of the hadron structure in this energy
regime is necessary to describe the strong interactions as they
are sensitive to the pion cloud and provide a test for the QCD
inspired effective field theories based on the chiral symmetry. A
promising approach is offered by constituent-quark models (CQMs).
Modern CQMs can be constructed so as to include the relevant
properties of QCD in the nonperturbative regime, notably the
consequences of the spontaneous breaking of chiral symmetry
($\chi$SB).

\section{Electromagnetic form factors}

The internal structure of nucleon is determined in terms of
electromagnetic Dirac and Pauli from factors $F_1(Q^2)$ and
$F_2(Q^2)$ or equivalently in terms of the electric and magnetic
Sachs form factors $G_E(Q^2)$ and $G_M(Q^2)$ \cite{sach}. The
issue of determination of the form factors has been revisited in
the recent past with several new experiments measuring the form
factors with precision at MAMI \cite{mami} and JLAB \cite{jlab}.
It has been shown that the proton form factors determined from the
measurements of polarization transfer \cite{jlab} were in
significant disagreement with those obtained from the Rosenbluth
separation \cite{rosen1}. This inconsistency leads to a large
uncertainty in our knowledge of the proton electromagnetic form
factors and urge the necessity for the new parameterizations and
analysis \cite{fit}.

The most general form of the hadronic current for a spin
$\frac{1}{2}$-nucleon with internal structure is given as \be
\langle B| J^\mu_{{\rm had}}(0)|B \rangle = \bar{u}(p')
\left(\gamma^\mu F_1(Q^2)+ i{\sigma^{\mu\nu}\over 2M}q_\nu
F_2(Q^2)\right) u(p), \label{ff} \ee where $u(p)$ and $u(p')$ are
the 4-spinors of the nucleon in the initial and final states
respectively. The Dirac and Pauli form factors $F_1(Q^2)$ and
$F_2(Q^2)$ are the only two form factors allowed by relativistic
invariance. These form factors are normalized in such a way that
at $Q^2=0$, they reduces to electric charge and the anomalous
magnetic moment in units of the elementary charge and the nuclear
magneton $\mu_N$, for example, \bea F^p_1(0)= 1\,,\,~~~~ F^p_2(0)
= \kappa_p = 1.793 \,, \,~~~~ F^n_1(0) = 0\,,\,~~~~ F^n_2(0) =
\kappa_n = -1.913\,.\eea

In analogy with the non-relativistic physics, we can associate the
form factors with the Fourier transforms of the charge and
magnetization densities. However, the charge distribution
$\rho(\bf{r})$ has to be calculated by a 3-dimensional Fourier
transform of the form factor as function of $\bf{q}$, whereas the
form factors are generally functions of $Q^2=\bf{q}{^2}-\omega^2$.
It would be important to mention here that there exists a special
Lorentz frame, the Breit or brick-wall frame, in which the energy
of the (space-like) virtual photon vanishes. This can be realized
by choosing $\bf{p}_1= -\frac{1}{2}\bf{q}$ and $\bf{p}_2=
+\frac{1}{2}\bf{q}$ leading to
$E_1=E_2 \,,$ 
$\omega=0$ and $Q^2=\bf{q}^2$. Thus, in the Breit frame, Eq.
(\ref{ff}) takes the following form \cite{sach} \be J_{\mu} =
\left( G_E(Q^2)\,, \iota
\frac{\bf{\sigma}\times\bf{q}}{2M}G_M(Q^2) \right)\,,
\label{eq}\ee where $G_E(Q^2)$ stands for the time-like component
of $J_{\mu}$ hence identified with the Fourier transform of the
electric charge distribution, whereas $G_M(Q^2)$ is interpreted as
the Fourier transform of the magnetization density. The Sachs form
factors $G_E$ and $G_M$ can be related to the Dirac and Pauli form
factors as \be G_E(Q^2)= F_1(Q^2)- \tau F_2(Q^2)\ ,~~~~~~~~~~
G_M(Q^2)= F_1(Q^2)+ F_2(Q^2)\,,\label{sach} \ee where
$\tau=\frac{Q^2}{4 M_N^2}$ is a measure of relativistic effects.

The  Fourier transform of the Sachs form factors can be expressed
as \be G_E ({\bf{q}}^2)=\int\rho({ \bf r}) e^{i{\bf q}\cdot{\bf
r}} d^3 {\bf r} ~~~= \int\rho({\bf r})d^3 {\bf r}-
{\frac{\bf{q}^2}{6}}\int \rho({\bf{r}}){\bf r}^2 d^3 {\bf r} +\
...\ , \ee where the first integral yields the total charge in
units of $e$, i.e., 1 for the proton and $0$ for the neutron, and
the second integral defines the square of the electric mean square
charge radius.

\section{charge radii of the nucleon}

The mean square charge radius of a given baryon ($r^2_B$) is one
of the important low energy characteristic giving its possible
``size'' and its precise determination give information about the
internal structure of the baryons. In general, $r^2_B$, which is a
scalar under spatial rotation is defined as $ r^2_B =  \int d^3 r
\rho(\bf r) r^2,$ where $\rho(\bf r)$ is the charge density. A
charge radius is the first nontrivial moment of a Coulomb monopole
$G_{C0}(q^2)$ transition amplitude.

In the recent past, with the advent of new facilities at JLAB,
SELEX Collaborations, the baryons charge radii are being
investigated. The results are available for the charge radii of
$p$, $n$, and very recently for the strange baryon ${\Sigma^-}$. For the case of proton we have $r_p = 0.877 \pm
0.007$ fm$^2$ ( $r^2_p = 0.779 \pm 0.025$ fm$^2$
\cite{rosenfelder}), for neutron we have $r^2_n = -0.1161 \pm
0.0022$ fm$^2$, \cite{pdg}, and for the case of $\Sigma^-$ we have
$r^2_{\Sigma^-} = 0.61 \pm 0.21$ fm$^2$ \cite{sigma-}. The
measurement of the $\Sigma^-$ charge radii is important as it
particularly suggests the possibility of measuring the charge
radii of other long-lived strange baryons such as $\Lambda$,
$\Sigma$, and $\Xi$ in the near future.

In the general parameterization (GP) method  \cite{morp}, the
charge radii operator can be expressed in terms of the sum of
one-, two-, and three-quark contributions \be \label{rad} {r}^2_B
= A \sum_{i=1}^3 e_i {\bf 1} + B\sum_{i \ne j}^3 e_i \,
{\bf{\sigma}_i} \cdot \bf{\sigma}_j  + C\sum_{i \ne j \ne k }^3
e_k \, \bf{\sigma}_i \cdot \bf{\sigma}_j \,,\ee where $e_i$ and
$\bf{\sigma}_i$ are the charge and spin of the i-th quark. The
constants $A$, $B$, and $C$ can be determined from the
experimental observations on charge radii and quadrupole moments
of the baryons.

The charge radii for the octet baryons can be calculated by
evaluating matrix elements of the operator in Eq. (\ref{rad})
between spin-flavor wave functions $|B \rangle $ as $\langle B
|r^2| B\rangle$. It is straightforward to verify that, for the
octet baryons, the operators involving two- and three-quark terms
in Eq. (\ref{rad}) can be simplified as \bea \sum_{i \neq j} e_i(
\bf{\sigma_i} \cdot \bf{\sigma_j}) = 2 \bf{J} \cdot \sum_{i} e_i
\bf{\sigma_i} - 3 \sum_{i} e_i \,, ~~~~~~~~~~~~ \sum_{i \neq j
\neq k} e_i(\bf{\sigma_j} \cdot \bf{\sigma_k})= - 3 \sum_i e_i -
\sum_{i \neq j} e_i( \bf{\sigma_i} \cdot \bf{\sigma_j})
\,.\label{eisjsk} \eea Using the expectation value of operator $2
J \cdot \sum_{i} e_i \bf{\sigma_i}$ between the baryon
wavefunctions $|B \rangle$ in the initial and final state baryons,
the operators in Eq. (\ref{eisjsk}) become \bea \sum_{{i \neq j}}
e_i(\bf{\sigma_i} \cdot \bf{\sigma_j}) = 3 \sum_{i} e_i
\bf{\sigma_{iz}} - 3 \sum_{i} e_i \,, ~~~~~~~~ \sum_{i\neq j \neq
k} e_i(\bf{\sigma_j} \cdot \bf{\sigma_k}) = - 3 \sum_{i} e_i
\bf{\sigma_{iz}}  \,,\eea The charge radii in Eq. (\ref{rad}) for
the octet baryons can now be expressed as \be {r^2_{B}} =
({{\mathrm A}}- 3{\mathrm B})\sum_i e_i + 3( {\mathrm B} -
{\mathrm C}) \sum_i e_i \sigma_{i z}\,, \label{r1/2} \ee It is
clear from this expression that the determination of charge radii
reduces to the calculation of the flavor and spin structure of
given octet baryon ($\widehat e_{i} \equiv \langle B| \sum_i
e_{i}|B \rangle \,,$ and $ \widehat {e_{i}\sigma_{iz}} \equiv
\langle B| \sum_i {e_i} {\sigma_{iz}} |B \rangle $). Here $|B
\rangle$ is the baryon wave function and  $(\sum_i e_{i})$ and
$(\sum_i e_i \sigma_{iz})$ are the charge and spin operators
defined as
\be
\sum_i e_i = \sum_{q=u,d,s} n^B_{q}q + \sum_{{\bar q} = {\bar u},
{\bar d}, {\bar s}} n^B_{{\bar q}} {\bar q} = n^B_{u}u + n^B_{d}d
+ n^B_{s}s + n^B_{\bar u}{\bar u} + n^B_{ \bar d}{\bar d} +
n^B_{\bar s}{\bar s} \label{numei} \,, \ee \be \sum_i {e_i}
{\sigma_{iz}} = \sum_{q=u,d,s} (n^B_{q_{+}}q_{+} +
n^B_{q_{-}}q_{-}) = n^B_{u_{+}}u_{+} + n^B_{u_{-}}u_{-} +
n^B_{d_{+}}d_{+} + n^B_{d_{-}}d_{-} + n^B_{s_{+}}s_{+} +
n^B_{s_{-}}s_{-} \label{numeisi} \,, \ee where $n_q^B$ $(n_{\bar
q}^B)$ is the number of quarks with charge $q$ (${\bar q}$),
$n^B_{q_{+}}$($n^B_{q_{-}}$) being the number of quarks with spin
$q_{+}$($q_{-}$) quarks.

Using the ${\rm SU}(6)$ spin-flavor symmetry of the wave functions
in the naive quark model (NQM), the charge radii of proton and
neutron becomes \be r^2_p = ({\mathrm A}-3 {\mathrm B})(2 u + d)+
3({\mathrm B} - {\mathrm C}) \left( \frac{4}{3}u_+ - \frac{1}{3}
d_+ \right) = {\mathrm A} - 3 {\mathrm C}\,, \label{crp} \ee  \be
r^2_n = ( {\mathrm A} - 3 {\mathrm B})(u +2 d)+ 3( {\mathrm B}-
{\mathrm C})\left( - \frac{1}{3}u_+ + \frac{4}{3} d_+ \right) =
-2{\mathrm B} + 2{\mathrm C} \,. \label{crn} \ee The naive quark
model (NQM) calculations show that the results are in disagreement
with the available experimental data. In this context, it
therefore becomes desirable to extend this model to understand the
role played by chiral symmetry breaking.

\section{Chiral symmetry breaking}

The global symmetry which arises in the QCD lagrangian, if we
neglect the small quark masses and consider the light quarks as
massless particles, is the chiral symmetry of SU(3)$_L$ $\times$
SU(3)$_R$ group. Since the spectrum of the hadrons in the known
sector, does not display parity doublets, we believe that the
chiral symmetry is spontaneously broken around a scale of 1 GeV as
\be SU(3)_L \times SU(3)_R \to SU(3)_{L+R} \to SU(3)_{V}\,. \ee As
a consequence there exist a set of massless particles called the
Goldstone bosons (GBs) which are further identified with the
observed ($\pi$, $K$, $\eta$). Although these are massive but are
interpreted as the GBs of the spontaneously broken chiral symmetry
as their masses are small compared to the nucleon mass. The QCD
Lagrangian is also invariant under the axial $U(1)$ symmetry, this
breaking symmetry picks the $\eta'$ as the ninth GBs.

If QCD leads to quark confinement, the mass parameters $m_q$, are
not directly observable quantities. However, they can be
determined in terms of observable hadronic masses through current
algebra methods. These quark masses are called current quark
masses in order to distinguish them from constituent quark masses.
Constituent quark masses, also called effective quark masses, are
parameters used in phenomenological quark models of hadronic
structures and are in general larger than the current quark
masses.

\section{chiral constituent quark model}

One of the most successful model in the nonperturbative regime of
QCD which incorporates $\chi$SB is the chiral constituent quark
model ($\chi$CQM) \cite{hd}. The basic process in the $\chi$CQM is
the emission of a GB by a constituent quark which further splits
into a $q \bar q$ pair as $ q_{\pm} \rightarrow {\rm GB}^{0} +
q^{'}_{\mp} \rightarrow (q \bar q^{'}) +q_{\mp}^{'}\,,
\label{basic} $ where $q \bar q^{'} +q^{'}$ constitute the ``quark
sea'' \cite{cheng,song,johan}. The effective Lagrangian describing
interaction between quarks and a nonet of GBs is ${\cal L} = g_8
{\bar q} \Phi q \,,$ \bea q =\left( \ba{c} u \\ d\\ s \ea \right),&
~~~~~& \Phi = \left( \ba{ccc} \frac{\pi^o}{\sqrt 2}
+\beta\frac{\eta}{\sqrt 6}+\zeta\frac{\eta^{'}}{\sqrt 3} & \pi^+ &
\alpha K^+   \\ \pi^- & -\frac{\pi^o}{\sqrt 2} +\beta
\frac{\eta}{\sqrt 6} + \zeta\frac{\eta^{'}}{\sqrt 3}  &  \alpha
K^o
\\ \alpha K^-  & \alpha \bar{K}^o & -\beta \frac{2\eta}{\sqrt 6}
+\zeta\frac{\eta^{'}}{\sqrt 3} \ea \right), \eea where $g_8$ and
$\zeta$ are the coupling constants for the singlet and octet GBs.
SU(3) symmetry breaking is introduced by considering $M_s >
M_{u,d}$ as well as by considering the masses of GBs to be
nondegenerate $(M_{K,\eta} > M_{\pi}$ and $M_{\eta^{'}}
> M_{K,\eta})$ {\cite{hd,cheng,song,johan}. The parameter
$a(=|g_8|^2$) denotes the transition probability of chiral
fluctuation of the splittings $u(d) \rightarrow d(u) +
\pi^{+(-)}$, whereas $\alpha^2 a$, $\beta^2 a$ and $\zeta^2 a$
respectively denote the probabilities of transitions of~ $u(d)
\rightarrow s  + K^{-(o)}$, $u(d,s) \rightarrow u(d,s) + \eta$,
 and $u(d,s) \rightarrow u(d,s) + \eta^{'}$.

The charge radii operator in Eq. (\ref{r1/2}), involves the
knowledge of spin and flavor structure of baryons in $\chi$CQM. A
redistribution of flavor and spin takes place among the ``sea
quarks'' in the interior of hadron due to the fluctuation process
and chiral symmetry breaking  in the $\chi$CQM. The flavor and
spin content can now be calculated by substituting for every
constituent quark by \bea q \to   P_q q + |\psi(q)|^2 \,,
~~~~~~~~~~ q_{\pm} \to P_q q_{\pm}+ |\psi(q_{\pm})|^2 \,,
\label{spin} \eea where $P_q = 1 - \sum P_q$ is the transition
probability of no emission of GB from any of the $q$ quark,
$|\psi(q)|^2$ is the transition probability of the $q$ quark, and
$|\psi(q_{\pm})|^2$ is the probability of transforming a $q_{\pm}$
quark \cite{hd}.

\section{Results and Discussion}

Recently it has been observed that the baryon wavefunction is
modified due to the minimal configuration mixing generated by the
chromodynamic spin-spin forces and it improves the low energy
dynamics of hadrons. Configuration mixing and details of the spin,
isospin and spatial parts of the wavefunction, can be found in
Refs. \cite{dgg,isgur,yaoubook}. Using the mixed wavefunction in
GP method, the charge radii for the proton and neutron can now be
expressed as \be r^2_p = ({\mathrm A}-3 {\mathrm B})(2u + d)+
3({\mathrm B} - {\mathrm C }) \bigg[ \cos^2 \phi \left(
\frac{4}{3}u_+ - \frac{1}{3} d_+ \right) + \sin^2 \phi \left(
\frac{2}{3}u_+ + \frac{1}{3} d_+ \right) \bigg ]\,, \label{crp1}
\ee \be r^2_n = ( {\mathrm A} - 3{\mathrm B})(u + 2d)+ 3( {\mathrm
B}- {\mathrm C})\bigg[ \cos^2 \phi \left(- \frac{1}{3}u_+ +
\frac{4}{3} d_+ \right) + \sin^2 \phi \left(\frac{1}{3}u_+ +
\frac{2}{3} d_+ \right) \bigg]\,. \label{crn1} \ee The modified
charge radii in the $\chi$CQM, after the modified quark content
from  Eq. (\ref{spin}), is expressed as \be r^2_p= (A- 3B)(1- a -2
a \alpha^2)+ 3(B- C)\left( \cos^2 \phi \left(1 -\frac{a}{3} (4+
2\alpha^2+ \beta^2+ 2\zeta^2)\right) + \sin^2 \phi \left(
\frac{1}{3} -\frac{a}{9}(6+ \beta^2+ 2\zeta^2) \right) \right)\,,
\ee \be r^2_n= (A-3B)(1- a \alpha^2)+ 3(B- C)\left( \cos^2 \phi
\left(-\frac{2}{3}+ \frac{a}{9} (3 + 9\alpha^2+ 2\beta^2+
4\zeta^2)\right) - \sin^2 \phi \frac{a}{3}\left( 1- \alpha^2
\right) \right)\,.\ee

The charge radii of the other octet baryons can be calculated in a
similar manner. For the $\chi$CQM parameters, we have used the
same set of symmetry breaking parameters as in Ref. \cite{hd} $a =
0.12\,, \alpha = 0.45\,, \beta = 0.45\,, \zeta = -0.15 \,.$ In
addition to the $\chi$CQM parameters, the GP parameters
corresponding to the one-, two- and three-quark contributions
($\mathrm A$, $\mathrm B$, and $\mathrm C$, respectively), have to
be fitted to the numerical values of charge radii of proton,
neutron and their quadrupole moment. The best fit values obtained
are ${\mathrm A}$ = 0.921, ${\mathrm B}$ = 0.105, and ${\mathrm
C}$ = 0.032, respectively. We have presented the $\chi$CQM$_{{\rm
config}}$ results for the charge radii of  octet baryons in Table
\ref{chargeradii1/2}. For the sake of comparison, we have also
presented the results for NQM and $\chi$CQM without configuration
mixing. A cursory look at the results reveal that the NQM and
$\chi$CQM predictions for the charge radii are on the higher side
as compared to the results with configuration mixing. As for the
case of other low energy hadronic matrix elements, the results
with configuration mixing in this case also are of the right order
of magnitude when compared with the available experimental data.
This gives a strong impetus to the spin-spin generated
configuration mixing in the baryon wavefunctions. This can be
particularly observed in the case of the charge radii of strange
baryon $\Sigma^-$ which matches well with experimental data. It is
interesting to observe that in this case there is no effect of
configuration mixing which can be easily understood when we look
into its detailed structure. Further, our model successfully
predicts the charge radii for the other octet baryons where there
is no experimental data available. Therefore, refinement of data
for the charge radii of other octet baryons would have important
implications for understanding the basic tenets of $\chi$CQM and
configuration mixing.

\section{ACKNOWLEDGMENTS}
H.D. would like to thank Department of Science and Technology,
Government of India and the organizers of CHIRAL2010:
International Workshop on Chiral Symmetry in Hadrons and Nuclei
for financial support.

\begin{table}
\begin{tabular}{|c|c|c|c|c|}\hline
Baryon & Data \cite{pdg, sigma-} & NQM & $\chi$CQM & {$\chi$CQM$_{{\rm
config}}$}
\\ \hline
$r^2_p$ &$r_p$ = 0.877$\pm$ 0.007 &0.809 &0.784 &0.769 \\ $r^2_n$
& $-$0.1161 $\pm$ 0.0022&$-$0.130 &$-$0.131 & $-$0.116 \\
$r^2_{\Sigma^+}$ &...&0.809 & 0.784 &0.769\\ $r^2_{\Sigma^-}$
&0.61 $\pm$0.21 &0.679 &0.675 &0.675 \\
$r^2_{\Sigma^0}$
&...&$-$0.065 & $-$0.054 & $-$0.047 \\ $r^2_{\Xi^0}$ &...&$-$0.130
& $-$0.131  & $-$0.116
\\ $r^2_{\Xi^-}$ &...& 0.679 & 0.675 & 0.675 \\ $r^2_{\Lambda}$
&...&$-$0.065 &$-$0.069 &$-$0.062 \\ \hline
\end{tabular} \caption{Charge radii of the octet baryons. }\label{chargeradii1/2}
\end{table}


\begin{thebibliography}{}



\bibitem{sach} R.G. Sachs, Phys. Rev. {\bf 126}, 2256 (1962).

\bibitem{mami} D.I. Glazier {\it et al.}, Eur. Phys. J. {\bf A 24},
101 (2005); D. Rohe, Eur. Phys. J. {\bf A 28} S1, 29 (2006).

\bibitem{jlab} M.K. Jones {\it et al.} (Jefferson Lab Hall A
Collaboration), Phys. Rev. Lett. {\bf 84}, 1398 (2000); O. Gayou
{\it et al.}, Phys. Rev. {\bf C 64}, 038202 (2001); O. Gayou {\it
et al.}, Phys. Rev. Lett. {\bf 88}, 092301 (2002); V. Punjabi {\it
et al.}, Phys. Rev. {\bf C 71}, 055202 (2005); C.B. Crawford {\it
et al.}, Phys. Rev. Lett. {\bf 98}, 052301 (2007).


\bibitem{rosen1} I.A. Qattan {\it et al.}, Phys. Rev. Lett. {\bf
94}, 142301 (2005).

\bibitem{fit} W.M. Alberico, S.M. Bilenky, C. Giunti, and K.M.
Graczyk, Phys. Rev. C {\bf 79}, 065204 (2009).

\bibitem{rosenfelder} R. Rosenfelder, Phys. Lett. B {\bf 479}, 381
(2000).

\bibitem{pdg} C. Amsler {\it et al.} (Particle Data Group), Phys.
Lett. B, {\bf 667} 1 (2008).

\bibitem{sigma-} I. Eschrich {\it et al.} (SELEX Collaboration),
Phys. Lett. B, {\bf 522}, 233 (2001).

\bibitem{morp} G. Morpurgo, Phys. Rev. D {\bf 40}, 2997 (1989).

\bibitem{hd} H. Dahiya and M. Gupta, Phys. Rev. D {\bf 64}, 014013
(2001); {\bf 66}, 051501(R) (2002); {\bf 67}, 114015 (2003); {\bf
67}, 074001 (2003); {\bf 78}, 014001 (2008);  N. Sharma, H.
Dahiya, P.K. Chatley and M. Gupta, Phys. Rev. D {\bf 79}, 077503
(2009);  {\bf 81}, 073001 (2010); N. Sharma, H. Dahiya, and P.K.
Chatley Eur. Phys. J. A {\bf 44}, 125 (2010); N. Sharma and H.
Dahiya, Phys. Rev. D {\bf 81}, 114003 (2010).


\bibitem{cheng} T.P. Cheng and Ling Fong Li, Phys. Rev. Lett. {\bf
74}, 2872 (1995); Phys. Rev. D {\bf 57}, 344 (1998).

\bibitem{song} X. Song, J.S. McCarthy and H.J. Weber, Phys. Rev.
D {\bf 55}, 2624 (1997); X. Song, Phys. Rev. D {\bf 57}, 4114
(1998).

\bibitem{johan} J. Linde, T. Ohlsson and H. Snellman, Phys. Rev. D
{\bf 57}, 452 (1998); {\bf 57} 5916 (1998).

\bibitem{dgg} A. De Rujula, H. Georgi, and S.L. Glashow, Phys. Rev.
D {\bf 12}, 147 (1975).

\bibitem{isgur} N. Isgur and G. Karl,  Phys. Rev. D {\bf 21}, 3175
(1980); P. Geiger and N. Isgur,  Phys. Rev. D {\bf 55}, 299
(1997).

\bibitem{yaoubook}  A. Le Yaouanc {\it et al.}, {\it Hadron
Transitions in the Quark Model}, Gordon and Breach (1988).


\end{thebibliography}
\end{document}